\documentclass{article}
\usepackage{amsfonts}
\usepackage{spconf,amsmath,graphicx}
\usepackage[colorlinks,linkcolor=blue]{hyperref}

\usepackage{caption}
\usepackage{pifont}
\usepackage{cite}
\usepackage{amssymb}
\usepackage{algorithmic}
\usepackage{algorithm}
\usepackage{textcomp}
\usepackage{marvosym}
\usepackage{float} 
\usepackage{pifont}

\title{MS-UNet-v2: Adaptive Denoising Method and Training Strategy for Medical Image Segmentation with Small Training Data}
\name{Haoyuan Chen, 
Yufei Han,
Pin Xu,
Yanyi Li,
Kuan Li*,
Jianping Yin}
\address{School of Cyberspace Security, Dongguan University of Technology, China}
\begin{document}
%
\maketitle

\begin{abstract}
Models based on U-like structures have improved the performance of medical image segmentation.
However, the single-layer decoder structure of U-Net is too "thin" to exploit enough information, 
resulting in large semantic differences between the encoder and decoder parts.
Things get worse if the number of training sets of data is not sufficiently large, which is common in medical image processing tasks where annotated data are more difficult to obtain than other tasks.
Based on this observation, we propose a novel U-Net model named MS-UNet for the medical image segmentation task in this study.
Instead of the single-layer U-Net decoder structure used in Swin-UNet and TransUnet, we specifically design a multi-scale nested decoder based on the Swin Transformer for U-Net. 
The proposed multi-scale nested decoder structure allows the feature mapping between the decoder and encoder to be semantically closer, thus enabling the network to learn more detailed features.
In addition, we propose a novel edge loss and a plug-and-play fine-tuning Denoising module, which not only effectively improves the segmentation performance of MS-UNet, but could also be applied to other models individually.
Experimental results show that MS-UNet could effectively improve the network performance with more efficient feature learning capability and exhibit more advanced performance, especially in the extreme case with a small amount of training data, and the proposed Edge loss and Denoising module could significantly enhance the segmentation performance of MS-UNet.
\end{abstract}
\begin{keywords}
Medical Image Segmentation, U-Net, Swin Transformer, Multi-scale Nested Decoder, Edge Loss, Plug-and-play Fine-tuning Denoising Module
\end{keywords}

\renewcommand{\dblfloatpagefraction}{.9}

\section{Introduction}
\label{sec:introduction}
Medical image segmentation plays a crucial role in computer-aided diagnosis and intelligent medicine, attracting significant interest in the medical image community.
CNN-based networks, such as U-Net\cite{UNet}, and their variants have been dominant in this field, showcasing notable performance improvements.
The symmetric encoder-decoder structure of U-Net, coupled with skip connections, allows for the extraction of rich image information while maintaining a relatively lightweight design.
Building upon U-Net's success, novel U-like models, including UNet++\cite{UNet++}, Res-UNet\cite{ResUNet}, and UNet3+\cite{UNet3+}, have been proposed and achieved excellent performances in medical image segmentation tasks.

However, CNN-based networks have gradually faced limitations in recent years.
The inherent inductive bias of CNN networks, which restrict each receptive field to a fixed-size window, hampers the establishment of long-range pixel dependencies.
In contrast, the Transformer architecture, originally designed for NLP tasks\cite{BERT}, has gained significant attention in the computer vision (CV) community.
Transformers, powered by the Multi-head Self Attention mechanism\cite{Transformer}, effectively establish global connections between tokens in sequences.
Vision Transformer (ViT)\cite{VisionTransformer}, the first pure transformer-based model, has revolutionized various computer vision tasks by processing 2D image patches with positional embeddings.
Furthermore, the hierarchical Swin Transformer, introduced in\cite{SwinTransformer}, reduces the computational complexity compared to\cite{VisionTransformer} and incorporates the local nature of CNNs.
This localization enables Transformers to handle pixel-level prediction tasks efficiently.
Motivated by the successes of Transformers and their combination with CNN architectures, we propose a novel Transformer-based U-like architecture named MS-UNet for 2D medical image segmentation.

Unlike the single-layer decoder structure in Swin-Unet\cite{SwinUnet}, inspired by\cite{UNet++, DoubleU-Net, DLA}, MS-UNet employs a hierarchical multi-scale nested decoder design.
This novel architecture allows MS-UNet to learn semantic information in the feature map from a more multi-dimensional perspective.
By overcoming the limitations of the single-layer U-shaped structure, MS-UNet offers improved information extraction and addresses the semantic differences between the encoder and decoder components.
This advantage is particularly crucial in medical image processing tasks, where annotated data is often limited.
Through the fusion of Transformer and U-like structures, MS-UNet represents a promising approach to enhance the performance of 2D medical image segmentation.
It leverages the power of Transformers while addressing the drawbacks of traditional CNN networks, especially in scenarios with limited training data.

\begin{figure*}
  \centering
  \includegraphics[width=0.9\textwidth]{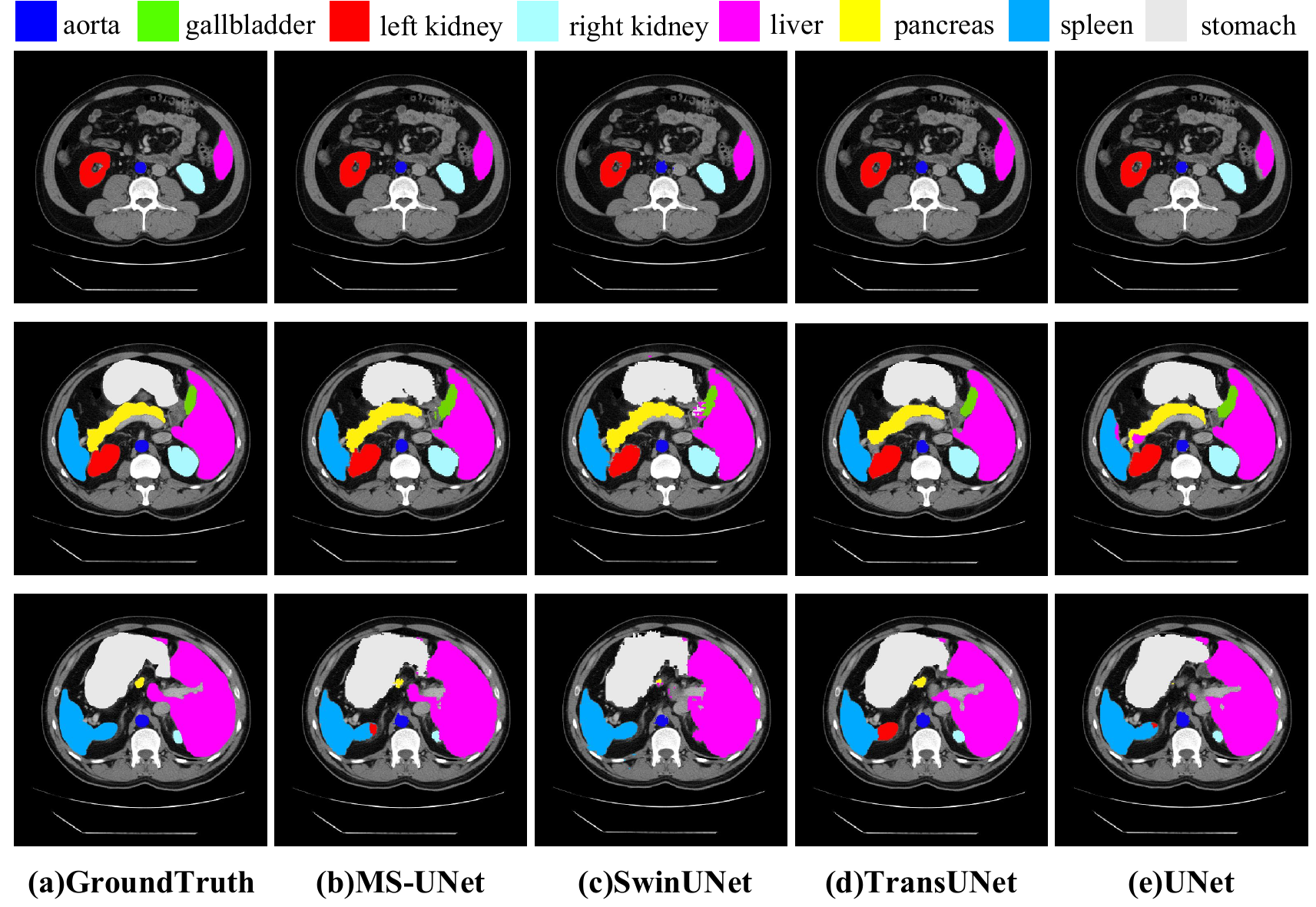}
\caption{Visualization of segmentation effects of different models on the Synapse multi-organ CT dataset.}
\label{fig:Visualization of segmentation1}
\end{figure*}

While network architecture plays a crucial role in improving segmentation performance, it is equally important to consider the impact of the loss function in medical image segmentation\cite{V-Net, Focal-Loss}.
Studies have shown that modifying the loss function could significantly enhance segmentation performance\cite{DE-Net}.
In recent years, researchers have explored the integration of edge-aware modules and corresponding loss functions to improve segmentation performance\cite{CM-MLP, PraNet, Ambiguous-Boundary, EAA-Net}.
For example, Hatamizadeh et al. introduced an edge-aware loss function that supervises an additional edge module, improving the network's sensitivity to edge information during training\cite{Boundary-Aware-Network}.
Similarly, Kuang et al. proposed an edge branch with an associated edge loss to enhance edge segmentation performance\cite{BEA-SegNet}.
However, these approaches often suffer from increased model complexity and training costs.
To address these challenges, our study proposes a novel Edge Loss function tailored for medical image segmentation.
The Edge Loss increases the network's sensitivity to target boundaries during training, resulting in finer segmentation edges.
Importantly, our Edge Loss achieves these improvements without introducing additional computational complexity or extending training time.
In previous research, similar approaches have been successful in natural image segmentation tasks.
Han et al. proposed a class-aware edge loss module that reduces segmentation errors in edge regions without affecting inference speed\cite{Edgenet}.
In the context of medical image segmentation, Gu et al. introduced the entirety-center-edge (ECE) loss function, which enhances boundary details without compromising overall performance\cite{DE-Net}.
Building upon these advancements, our Edge Loss function extends the benefits of edge-aware modules to the specific domain of medical image segmentation.
By emphasizing boundary information and optimizing boundary details, our proposed Edge Loss contributes to improved segmentation performance while maintaining computational efficiency and training time.
In this article, we present the results of our experiments with the proposed Edge Loss function in medical image segmentation tasks.
We demonstrate its effectiveness in enhancing the network's ability to capture precise boundary information, ultimately leading to more accurate and detailed segmentation results.

\begin{figure*}[htbp]
  \centering
  \includegraphics[width=\textwidth]{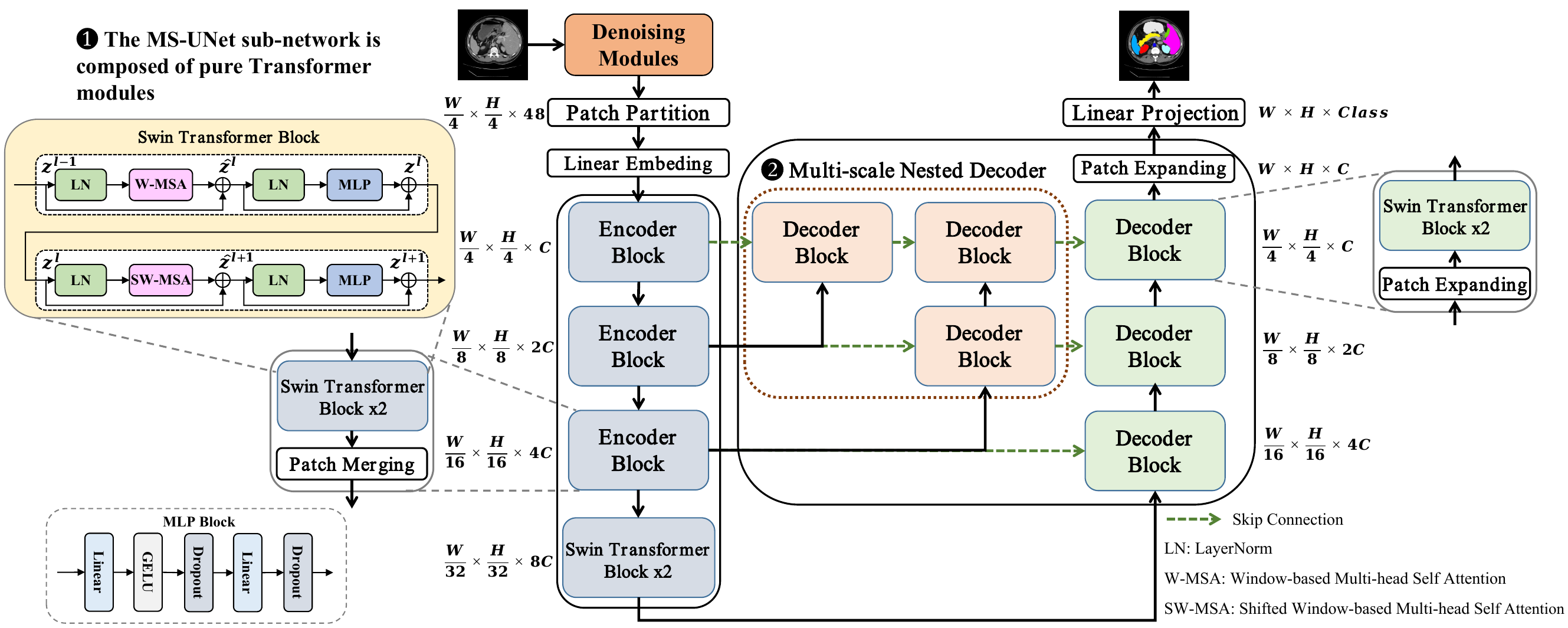}
\caption{The architecture of MS-UNet and plug-and-play fine-tuning denoising module. Our contributions: \ding{182} The MS-UNet sub-network is composed of pure Transformer modules; \ding{183} Multi-scale Nested Decoder.}
\label{fig:MS-UNet}
\end{figure*}

In the pursuit of improving medical image segmentation accuracy, numerous complex U-like family models have been proposed in recent years.
However, the scarcity of labeled medical image data poses a significant challenge, as data acquisition and annotation are costly\cite{Datasets}.
To mitigate this issue, researchers often resort to training models for multiple epochs using a limited dataset.
Moreover, medical images are susceptible to noise, which could obscure details and potentially hinder model recognition and analysis during training, leading to considerable performance losses\cite{Review-Denoising}.
Importantly, medical images often contain subtle details that are crucial for capturing essential medical features\cite{denoising-review}.
Inspired by the success of the "pre-training then fine-tuning" paradigm in computer vision tasks\cite{AIM}, and motivated by frozen-fine-tuning methods\cite{VPT} and trainable denoising techniques\cite{EDCNN, BEA-SegNet, Eformer}, we propose a novel fine-tuning trainable denoising module.
This module introduces a trainable denoising component while keeping the model backbone frozen, effectively reducing noise interference.
The key advantage of our fine-tuning trainable denoising module is its ability to prevent the network from learning incorrect features over multiple training epochs, thus preserving overall performance.
By incorporating the denoising module, our approach ensures that the model could capture important medical image features while effectively reducing the impact of noise.
Moreover, this module provides an opportunity for the model to learn and retain crucial but easily overlooked image details from the initial stages of training.
The proposed fine-tuning trainable denoising module addresses the challenges posed by limited medical image datasets and the presence of noise.
By leveraging the benefits of frozen-fine-tuning and trainable denoising techniques, our approach enhances the model's ability to accurately segment medical images while effectively reducing the interference caused by noise.
In the subsequent sections, we present experimental results that validate the efficacy of our method in improving the performance of medical image segmentation tasks.

In this paper, we present three key contributions to improve medical image segmentation performance.
First, we propose MS-UNet, a novel Transformer-based U-like architecture.
MS-UNet enhances the semantic hierarchy between the encoder and decoder, leading to improved stability and generalization.
By achieving a tighter semantic hierarchy, MS-UNet reduces the reliance on large-scale training data, which is particularly valuable in the medical image domain where labeled data is scarce.
To address the issue of segmentation edge quality, we introduce a novel edge loss and an efficient method for edge label extraction.
The proposed edge loss enhances the network's sensitivity to edge information without requiring additional network modules.
This approach avoids the added computational complexity and training time associated with performance improvement.
Moreover, we address the lack of edge label data in some datasets by devising an efficient edge label generation method using the Sobel operator.
This method accurately extracts edge-label data from ground truth images.
Additionally, we propose a fine-tuning trainable denoising module to mitigate the impact of image noise on model performance.
Unlike other denoising modules, our approach could be added externally to any well-trained model. By fine-tuning a few parameters, the module corrects the model's misinterpretation of image noise as significant feature information.
This fine-tuning process further improves the segmentation performance without the need for extensive retraining.
Experimental results demonstrate the effectiveness of our proposed methods.
MS-UNet, the edge loss, and the fine-tuning denoising module significantly enhance the network's performance, particularly in scenarios with limited training data.
These contributions highlight the efficacy of our approach in addressing challenges specific to medical image segmentation.
Our proposed methods exhibit outstanding performance and demonstrate more efficient feature learning capabilities, allowing it to be successfully applied to medical image segmentation tasks.

Our contributions are fivefold:
\begin{enumerate}
    \item A novel semantic segmentation framework with a multi-scale nested decoder, MS-UNet, is proposed to effectively improve the performance of the network with more efficient feature learning capability.
    More importantly, in the extreme case where only a very small amount of training data is available, MS-UNet achieves a huge improvement in segmentation results than other state-of-the-art models in the U-Net family.
    \item A novel edge loss is proposed to increase the sensitivity of the network to the boundary information without increased computational complexity and extra training time.
    It allows the network to output smoother edges, making the predictions closer to the ground truth.
    \item An efficient edge label generation module is proposed to obtain edge label data for datasets without edge labels.
    It ensures the edge loss could be applied to the majority of datasets.
    \item we propose a plug-and-play fine-tuning trainable denoising module, which could prevent the model from learning the wrong feature from image noise.
    Moreover, it could enable trained models to further improve segmentation performance with only a few fine-tuned parameters.
    \item We have successfully used MS-UNet with the edge loss and the denoising module for medical image processing tasks and have surpassed other state-of-the-art methods.
\end{enumerate}

The remainder of the paper is organized as follows.
Section II briefly reviews the related work.
In section III, we describe the details of our proposed MS-UNet, edge loss, and plug-and-play denoising module.
Section IV reports the results of MS-UNet on medical image segmentation tasks, as well as a series of ablation experiments.
We summarise our work in section V.

\section{Related Works}
\label{sec:related works}
We review some related works on medical image segmentation, edge loss function, and medical image denoising method below.

\subsection{Medical Image Segmentation Models}
The UNet model, which is based on convolutional neural networks (CNNs), has demonstrated outstanding performance in medical image segmentation.
Various extensions have been proposed to enhance its capabilities.
For instance, UNet++\cite{UNet++} introduces nested and dense connections to alleviate semantic differences within the network.
Res-UNet\cite{ResUNet} combines U-Net with ResNet and attention mechanisms, leading to improved performance in retinal vessel segmentation.

In recent years, the Transformer model has gained significant attention in computer vision tasks\cite{VisionTransformer}.
Its success has inspired researchers to explore its application in various domains, including medical image segmentation.
Chen et al. proposed the first Transformer-based medical image segmentation framework\cite{TransUNet}, where the CNN feature map is used as input to the Transformer encoder.
This approach achieved remarkable results in medical image segmentation tasks.
One notable development is the Swin Transformer\cite{SwinTransformer}, a hierarchical vision Transformer that combines the strengths of Transformers and CNNs.
By utilizing shifted windows, the Swin Transformer not only reduces computational complexity but also achieves state-of-the-art performance in computer vision tasks.
Building upon this, researchers introduced the concept of a pure Transformer network with a U-shaped Encoder-Decoder architecture for 2D medical image segmentation\cite{SwinTransformer}.
Additionally, the Dual Swin Transformer U-Net (DS-TransUNet)\cite{DS-TransUNet} was proposed, which incorporates the advantages of a hierarchical Swin Transformer into a standard U-shaped Encoder-Decoder architecture, effectively improving medical image segmentation performance.

Inspired by the advancements in U-like architectures and Transformer-based models, we propose MS-UNet, a simple but highly effective 2D medical image segmentation architecture based on the Swin Transformer.
MS-UNet distinguishes itself by replacing the single-layer decoder structure of Swin-UNet with a hierarchical multi-scale decoder structure.
This modification allows MS-UNet to learn semantic information from the feature map in a more comprehensive and multi-dimensional manner, leading to improved segmentation performance.

\begin{figure*}[!htbp]
  \centering
  \includegraphics[width=\textwidth]{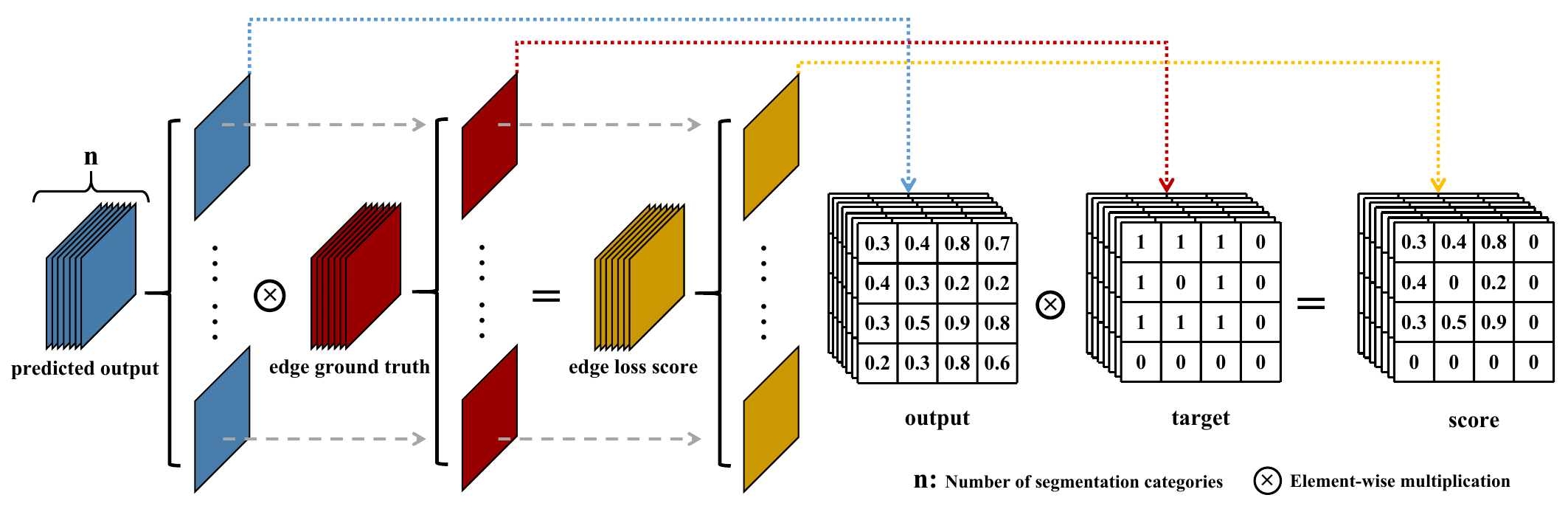}
\caption{The calculation process for the Edge Loss.}
\label{fig:Edge loss}
\end{figure*}
\begin{figure}
  \centering
  \includegraphics[width=\linewidth]{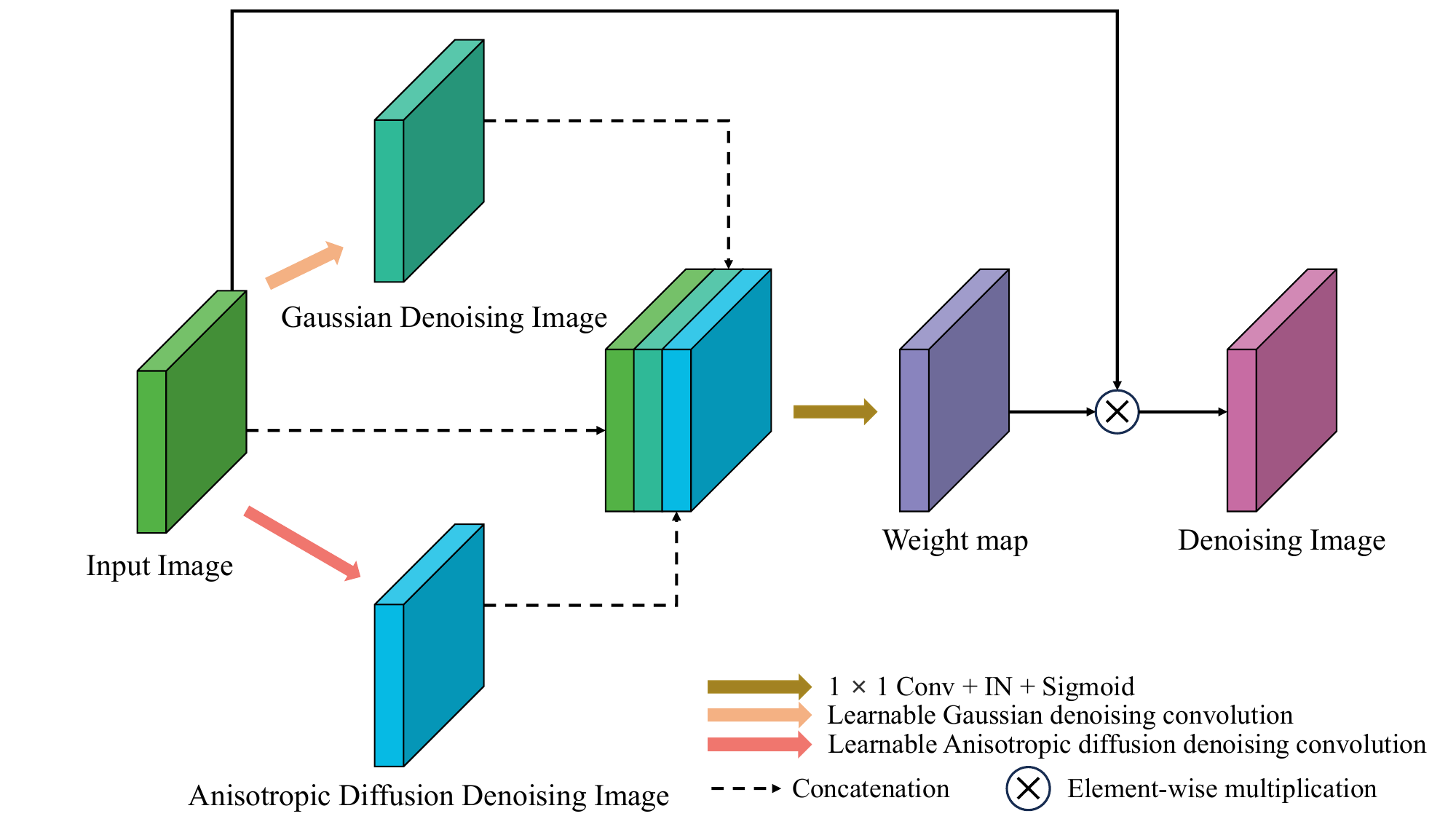}
\caption{The architecture of plug-and-play fine-tuning denoising module.}
\label{fig:denoising module}
\end{figure}

\subsection{Edge Loss}
In numerous studies aiming to enhance segmentation performance, researchers have recognized the crucial role of loss functions in image segmentation.
Particularly in medical image analysis, expert manual segmentation often relies on the accurate delineation of anatomical structure boundaries.
Building on this insight, Hatamizadeh et al. introduced boundary-aware CNNs for medical image segmentation, incorporating a dedicated edge branch and an edge-aware loss term to consider organ boundary information\cite{Boundary-Aware-Network}.
Similarly, Kuang et al. enhanced edge information perception in networks through an additional edge branching module with an edge loss, enabling the model to learn more effective edge segmentation features\cite{BEA-SegNet}.
Gu et al. designed an Edge Information Guiding Module (EIGM) to incorporate boundary information and proposed an overall-center-edge (ECE) loss function to further optimize boundary details by emphasizing boundary information\cite{DE-Net}.

While these studies have successfully improved the network's ability to learn edge information by incorporating additional edge branching modules and corresponding edge losses, they also introduce more parameters and increase the model's learning cost.
In this paper, we propose an edge loss that operates on the entire model, enabling the production of finer segmented edges without the need for additional edge branching modules.
Furthermore, we propose an algorithm for generating edge-labeled data using the Sobel operator, which efficiently produces edge labels for datasets lacking pre-existing edge annotations.

\subsection{Denoising methods}
In medical image processing tasks, the presence of noise could degrade image quality and potentially lead to incorrect predictions by the model.
Therefore, denoising is a crucial preprocessing technique in medical image processing\cite{Denoising-Techniques}.
Over the years, researchers have explored various denoising methods for medical images.
Liang et al. developed an edge enhancement module using a trainable Sobel convolution and constructed a densely connected model that integrates extracted edge information for end-to-end image denoising\cite{EDCNN}.
Building upon this, Luthra et al. combined the trainable Sobel-Feldman operator with Transformer and proposed the Eformer, an edge-enhanced Transformer-based architecture for medical image denoising\cite{Eformer}.
These methods have shown promising denoising results in medical image processing tasks; however, they often require significant computational resources.

Unlike natural images, medical images contain subtle details that are important for representing medical features.
The challenge lies in preserving these details during the denoising process.
Recently, the "pre-training then fine-tuning" paradigm has gained attention in the application of large models to computer vision tasks\cite{AIM} and Natural language processing (NLP) tasks\cite{Parameter-efficient}.
Motivated by this, we propose a novel fine-tuning trainable denoising module. This module reduces noise by introducing a trainable denoising component while keeping the model backbone frozen.
By adding a small number of parameters to the trained model, our approach effectively reduces medical image noise and prevents the model from incorrectly utilizing noise as important feature information, without sacrificing relevant medical features.

\begin{figure*}[htbp]
  \centering
  \includegraphics[width=0.9\textwidth]{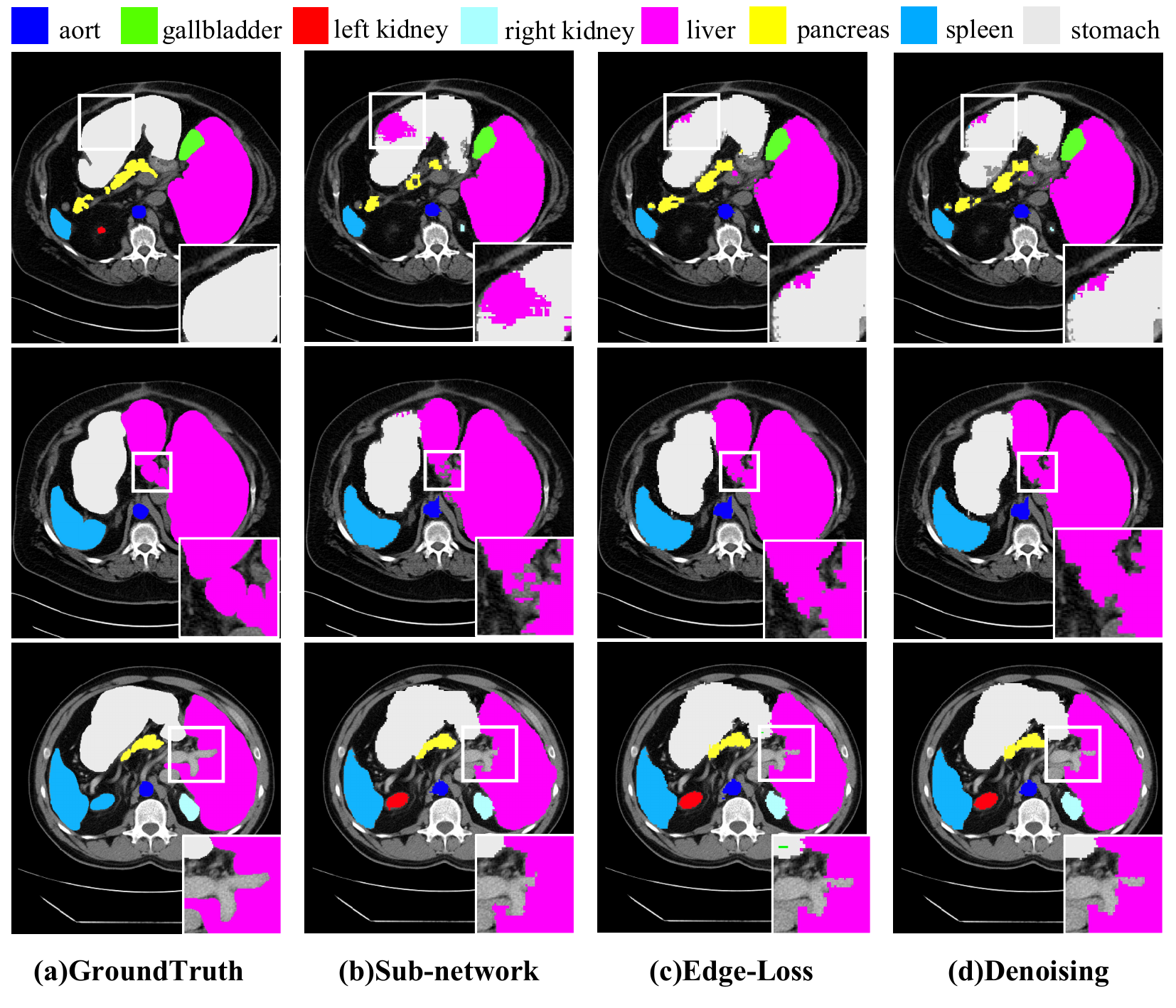}
\caption{Visualization of segmentation effects of different combinations of methods on the Synapse multi-organ CT dataset. (b) is the segmentation result with MS-UNet only. (c) is the segmentation result with MS-UNet and Edge-Loss. (d) is the segmentation result with the use of MS-UNet, Edge-loss, and plug-and-play fine-tuning denoising module.}
\label{fig:Visualization of segmentation2}
\end{figure*}

\section{Method}
\label{sec:method}
Fig.\ref{fig:MS-UNet} represents the MS-UNet that consists of a fine-tuning denoising module, MS-UNet sub-network, and loss function.
The fine-tuning denoising module consists of the trainable Gaussian block and the Anisotropic Diffusion block, which could preprocess the image and thus prevent the model from incorrectly using image noise as important feature information.
Difference from other U-like familiar architecture, the MS-UNet sub-network uses Swin Transformer instead of CNN as the backbone network.
It consists of an encoder step and a multi-scale nested decoder step, connected by skip connections.
Finally, the loss function consists of a total of three losses, namely the Dice-Similarity Coefficient (DSC) loss, the Hausdorff Distance (HD) loss, and our proposed edge loss.
Below, we would like to delve into each component of our MS-UNet in detail.

\subsection{MS-UNet Sub-Network}
The MS-UNet sub-network, an essential component of our methodology, combines the power of Swin Transformer and U-like architectures to revolutionize medical image segmentation.
It comprises an encoder step, a multi-scale nested decoder step, and skip connections that seamlessly connect the two.
In the encoder step, we partition the medical image into non-overlapping patches and create a raw-valued feature by concatenating the original pixel values.
This feature is then transformed through a series of Swin Transformer blocks and Patch Merging layers, capturing hierarchical representations and global context.
The multi-scale nested decoder step employs skip connections to upsample and decode the features obtained from each encoder layer independently.
This novel approach allows the decoder to learn semantic information from multiple dimensions, addressing the semantic differences between the encoder and decoder parts.
The resulting features are combined with the corresponding decoder input features, enhancing the network's stability and generalization.
Finally, we employ an upsampling Patch Expanding layer to restore the feature resolution and use a Linear Projection layer for pixel-level segmentation prediction.

The MS-UNet sub-network leverages the strengths of Swin Transformer and the U-like architecture to improve the stability, generalization, and segmentation performance of the model.
By integrating hierarchical representations, multi-dimensional feature learning, and skip connections, our methodology offers a promising approach for accurate medical image segmentation.

\subsubsection{Encoder Step}
The encoder branch of our proposed MS-UNet utilizes a hierarchical network architecture with Swin Transformer blocks as the backbone.
This design choice allows for an effective representation of learning from the input medical image.
Each Swin Transformer block plays a crucial role in this process, incorporating the Patch Merging layer to downsample the input information and increase the feature dimension.
By doing so, the network retains the essential information in the image while expanding the perceptual field.
This hierarchical approach enables the encoder branch to capture meaningful features at multiple scales, facilitating more comprehensive representation learning for subsequent decoding and segmentation tasks. 

\subsubsection{Mutil-Scale Nested Decoder and Skip Connection}
Similar to the encoder branch, the decoder branch of our MS-UNet also utilizes Swin Transformer as the backbone, accompanied by upsampling modules and Patch Expanding layers.
This symmetrical design ensures consistency and correspondence between the encoding and decoding stages of the network.
In contrast to traditional U-shaped models that rely on simple skip-connections to connect Encoder Blocks and Decode Blocks, we propose a novel approach: the multi-scale nested decoder branch.
This innovation replaces the conventional decoder branch, enhancing the network's ability to extract more effective information from the complex Transformer encoder output features.
The multi-scale nested blocks in our decoder branch perform two essential functions.
Firstly, they upsample the decoded features from each encoder independently, enabling the integration of information at different scales.
Secondly, they leverage skip connections to concatenate the resulting features with the corresponding decoder input features in the subsequent layer.
This fusion process aligns the semantic levels of the encoder and decoder feature maps, facilitating better information exchange and feature consistency.

Our multi-scale nested decoder structure not only enables effective feature fusion between different scales but also facilitates feature fusion between adjacent decoders.
This comprehensive utilization of skip connections ensures efficient information flow throughout the decoder network.
Importantly, the proposed multi-scale nested decoder structure empowers the network to learn the necessary information more efficiently, particularly in situations where limited training data is available.
This characteristic enhances the network's stability, generalization capability, and performance, making it suitable for medical image segmentation tasks even with small datasets.

\subsection{Loss Functions}
The loss functions we use in the training process are the Cross-Entropy loss function, the Dice-Similarity Coefficient loss function, and our proposed edge loss function.
The equation for this concept is:
\begin{equation}
\mathcal{L}_{total} = \omega_1\mathcal{L}_{CE} + \omega_2\mathcal{L}_{Dice} + \omega_3\mathcal{L}_{Edge},
\label{Loss}
\end{equation}
In the testing process, our evaluation metrics for the segmentation performance of the model consist of the Dice-Similarity Coefficient function and the Hausdorff Distance function.
In the following, we will describe the details of each loss function.

\subsubsection{Cross-Entropy Loss}
Cross-Entropy loss is the most popular loss function for image semantic segmentation tasks, which examines each pixel individually and compares the predictions for each pixel class with the label vector.
The equation for this concept is:
\begin{equation}
\ell_n = -\log\dfrac{\exp(x_n,y_n)}{{\sum_{c=1}^C}\exp(x_n,c)}\cdot{1},
\end{equation}
\begin{equation}
\mathcal{L}_{CE}(x,y) = {\sum_{n=1}^N}\dfrac{\ell_n}{N},
\end{equation}
where $C$ is the number of classes, $x$ is the input, $y$ is the grand true, and $N$ spans the minibatch dimension.

\subsubsection{Dice-Similarity Coefficient (DSC)}
The Dice coefficient is a metric function for calculating the similarity of sets. 
It is usually used to calculate the similarity between two samples, and its value ranges from [0, 1]. 
The equation for this concept is:
\begin{equation}
\mathcal{L}_{DSC} = 1 - \dfrac{|X \cap Y|}{|X| + |Y|},
\end{equation}
where X and Y are two sets. 
The set in the $|\cdot|$ represents the cardinality of the set, that is, the number of elements in the set. 
E.g. $|X|$ refers to the number of elements in set X. 
${\cap}$ is used to represent the intersection of two sets and means the elements that are common to both sets.

\subsubsection{Edge-Loss}
As our experiments will show, the model outputs have a heavily jagged edge compared to the grand true.
However, in medical image analysis, the boundaries of anatomical structures are often an important basis for manual segmentation by specialists.
It would lead to the loss of detailed information and reduce the usefulness of the predictions.
Therefore, we propose a novel edge loss to improve the sensitivity of the model to image boundary information so as to output more accurate segmentation results with smooth edges.
The equation for this concept is:
\begin{equation}
\ell_{edge} = \dfrac{2 \times sum(target * score)}{sum(target * target) + sum(score * score)},
\end{equation}
\begin{equation}
\mathcal{L}_{Edge} = 1 - \ell_{edge}
\end{equation}
where the target is the tensor of the edge grand trues, the score is the tensor of results calculated from the output and the edge grand trues, and $*$ represents the point-to-point multiplication between two tensors, $sum(\cdot)$ represents the sum of the elements within the tensor.
The detailed calculation process for the score is illustrated in Fig.\ref{fig:Edge loss}.

Different from other models, our proposed edge loss was not applied to the output of a particular edge branch prediction module, but directly to the final output of the overall model.
In other words, the edge loss would help us to effectively increase the sensitivity to edge information in the model without additional modules.
Therefore, there is no additional computational complexity and training time for the model for better performance.
More importantly, it could also be applied to any other model, enhancing the ability of models to learn from edge information.

\subsubsection{Hausdorff Distance (HD)}
Hausdorff distance is a measure that describes the similarity between two sets of points by calculating the distance between them. 
Suppose there are two sets of sets: $A=\lbrace{a^1,a^2,...,a^n}\rbrace$, $B=\lbrace{b^1,b^2,...,b^n}\rbrace$, the Hausdorff distance between these two sets of points are defined as:
\begin{equation}
H(A, B) = max(h(A, B), h(B, A)),
\end{equation}
\begin{equation}
h(A, B) = \mathop{max}\limits_{a \in A}\mathop{min}\limits_{b \in B}\parallel{a-b}\parallel,
\end{equation}
\begin{equation}
h(B, A) = \mathop{max}\limits_{b \in B}\mathop{min}\limits_{a \in A}\parallel{b-a}\parallel,
\end{equation}
where $\parallel{\cdot}\parallel$ represents the Euclidean distance between elements.

\begin{algorithm}[htbp]
    \renewcommand{\algorithmicrequire}{\textbf{Input:}}
    \renewcommand{\algorithmicensure}{\textbf{Output:}}
    \caption{Edge Label Generation Process.}\label{alg:sobel_edge}
    \begin{algorithmic}
        \REQUIRE
            the Ground Truth of training data $\mathbf{X}$, the number of label categories $\mathbf{n}$
        \ENSURE
            The Edge Ground Truth of training data $\mathbf{Y}$.
        \STATE $\mathbf{Y} \gets$ an zero matrix like $\mathbf{X}$
        \STATE $i \gets 1$
        \WHILE{$i <= \mathbf{n}$}
        \STATE $\hat{X} \gets \mathbf{X}$
            \IF{elements $x$ in $\hat{X}$ = i}
            \STATE $x \gets 1$
            \ELSE
            \STATE $x \gets 0$
            \ENDIF
        \STATE $\hat{X} \gets $Sobel$(\hat{X})$
            \IF{elements $x$ in $\hat{X}$ >= 0.2}
            \STATE $x \gets i$
            \ELSE
            \STATE $x \gets 0$
            \ENDIF
            \IF{elements $x$ in $\hat{X} \neq$ same elements in $\mathbf{X}$}
            \STATE $x \gets 0$
            \ENDIF
        \STATE $\mathbf{Y} \gets \mathbf{Y} + \hat{X}$
        \ENDWHILE
        \STATE $ $return$ \mathbf{Y}$
    \end{algorithmic}
\end{algorithm}

\subsection{Edge Label Generation Module}
In medical image segmentation tasks, acquiring labeled data that includes specific parts such as edges could be challenging due to the time and effort required for manual annotation by pathologists.
Furthermore, obtaining edge label data from existing labeled data could also be resource-consuming for researchers.
To address this issue, we propose an efficient edge label generation module utilizing the Sobel operator.
The module algorithm is outlined in algorithm\ref{alg:sobel_edge}.

In this algorithm, we begin by extracting the corresponding image tensor from the ground truth dataset.
Next, we assign a value of $1$ to the label elements and set the remaining elements to the background ($0$) based on the characteristics of the Sobel operator.
The tensor is then fed into the Sobel convolution operation, which calculates the edge output.
As the elements of the output tensor range between $0$ and $1$, we set elements greater than $0.1$ as edge labels, resulting in a preliminary edge label tensor.
Finally, we combine the preliminary edge label tensor with the corresponding image tensor using logical conjunction to obtain the final edge label data.

By incorporating this edge label generation module, we could efficiently and accurately generate edge label data from the ground truth images.
This approach significantly enhances the diversity of labeled data, which is crucial in medical image tasks that suffer from limited availability of labeled data.

\subsection{Plug-and-Play Fine-Tuning Denoising Module}
In medical imaging tasks, the presence of low image quality, small objects, and low contrast poses challenges for models in effectively learning medical features.
This vulnerability to noise could make it difficult for models to accurately perform tasks such as segmentation.
Moreover, the limited availability of labeled training data in fully supervised medical segmentation further exacerbates this problem, as models may erroneously learn incorrect features during training.
Complicating matters further is the fact that medical images often contain small details that may consist of only a few pixels but carry crucial medical information.
Balancing the reduction of noise while preserving these important but unknown medical features presents a significant challenge for medical image-denoising algorithms.
In light of recent advancements in the "pre-training then fine-tuning" paradigm in the computer vision community\cite{AIM}, we propose a plug-and-play fine-tuning denoising module.
This approach ensures that models learn a sufficient number of features before mitigating the impact of noise, as depicted in Fig.\ref{fig:denoising module}.
By incorporating this module into the overall framework, we aim to strike a balance between noise reduction and feature preservation, ultimately improving the performance of medical image-denoising algorithms.  

We first trained the MS-UNet sub-network using original images to ensure that the model could learn detailed features of images.
After that, we freeze the parameters of the trained MS-UNet sub-network except for Linear Projection and introduce the fine-tuning denoising module.
In Fine-tuning denoising module, we first perform the trainable Gaussian convolution and Anisotropic Diffusion convolution on the original input to obtain two denoising images respectively.
In Anisotropic Diffusion convolution, we use an eight-way diffusion operation with 30 rounds.
The four-way diffusion equation for this concept is:
\begin{equation}
\begin{split}
I_{t+1} = I_t + \lambda(&cN_{x,y}\nabla_N(I_t) + cS_{x,y}\nabla_S(I_t)\\
&+ cE_{x,y}\nabla_E(I_t) + cW_{x,y}\nabla_W(I_t)),
\end{split}
\end{equation}
where $I$ represents the image and $t$ represents the diffusion round.
$\nabla_N$, $\nabla_S$, $\nabla_E$ and $\nabla_W$ represent the bias in four directions for the pixel $(x,y)$.
\begin{equation}
\begin{split}
\nabla_N(I_{x,y}) &= (I_{x,y-1} - I_{x,y}),\\
\nabla_S(I_{x,y}) &= (I_{x,y+1} - I_{x,y}),\\
\nabla_E(I_{x,y}) &= (I_{x-1,y} - I_{x,y}),\\
\nabla_W(I_{x,y}) &= (I_{x+1,y} - I_{x,y}),
\end{split}
\end{equation}
$cN$, $cS$, $cE$ and $cW$ represent the thermal conductivity in four directions.
\begin{equation}
\begin{split}
cN_{x,y} &= \exp(-\|\omega_N\nabla_N(I)\|^2 / k^2),\\
cS_{x,y} &= \exp(-\|\omega_S\nabla_S(I)\|^2 / k^2),\\
cE_{x,y} &= \exp(-\|\omega_E\nabla_E(I)\|^2 / k^2),\\
cW_{x,y} &= \exp(-\|\omega_W\nabla_W(I)\|^2 / k^2),
\end{split}
\end{equation}
where $\omega_N$, $\omega_S$, $\omega_E$ and $\omega_W$ represent trainable weight parameters in four directions.

Subsequently, the two noise-reduced images obtained by the above operations are concatenated with the original image and then processed by a $1\times1$ convolutional layer with a stride of $1$.
The final batch normalization, ReLU activation, and Sigmoid function are performed to produce a noise-reduced weight map.
With the module, we could reduce the impact of image noise on performance by heating a few parameters during fine-tuning training.

\begin{figure*}[!htbp]
\begin{minipage}{0.48\textwidth}
  \centering
  \includegraphics[width=\linewidth]{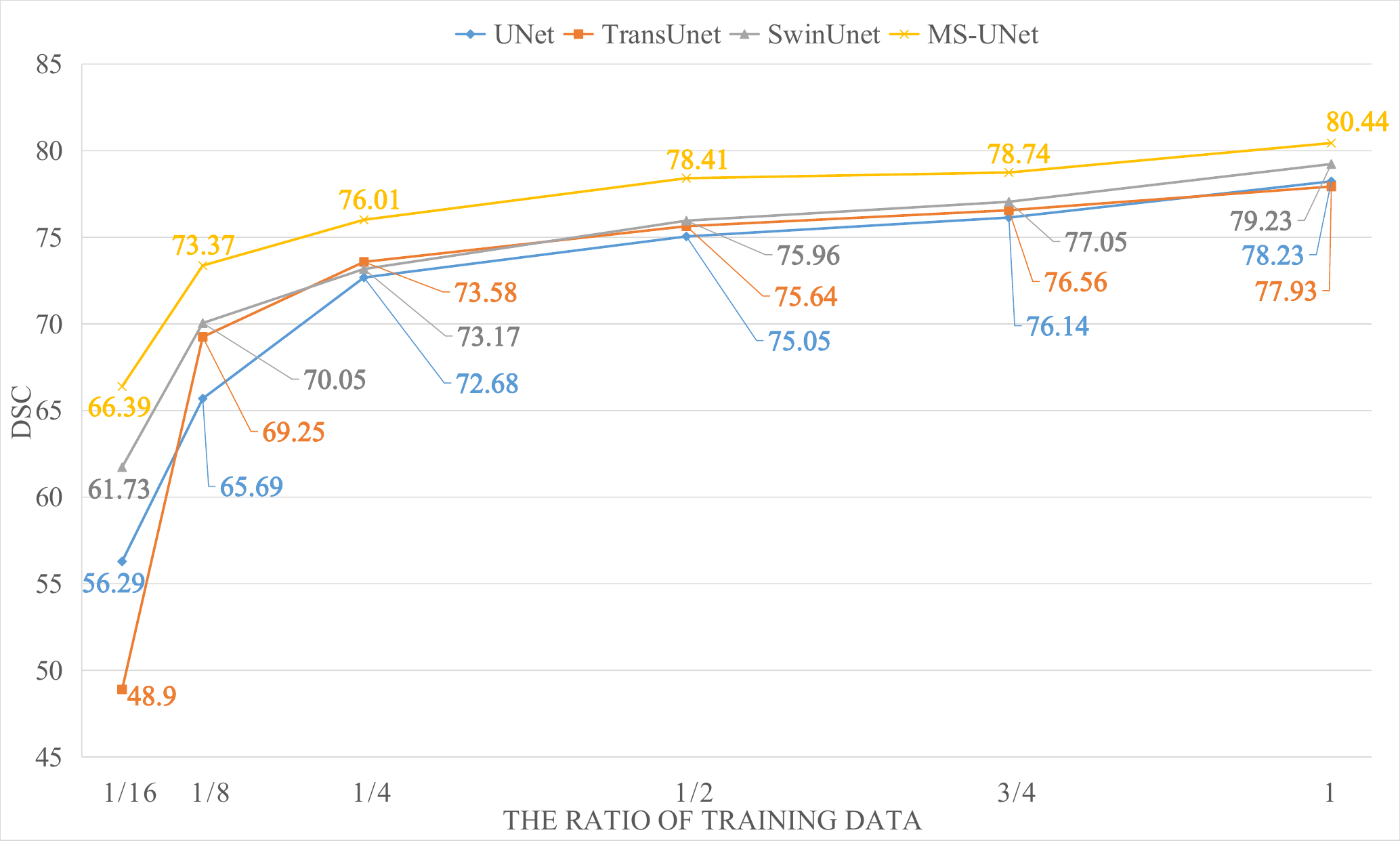}
  \centerline{(a) DSC on Synapse Dataset.}\medskip
\end{minipage}
\hfill
\begin{minipage}{0.48\textwidth}
  \centering
  \includegraphics[width=\linewidth]{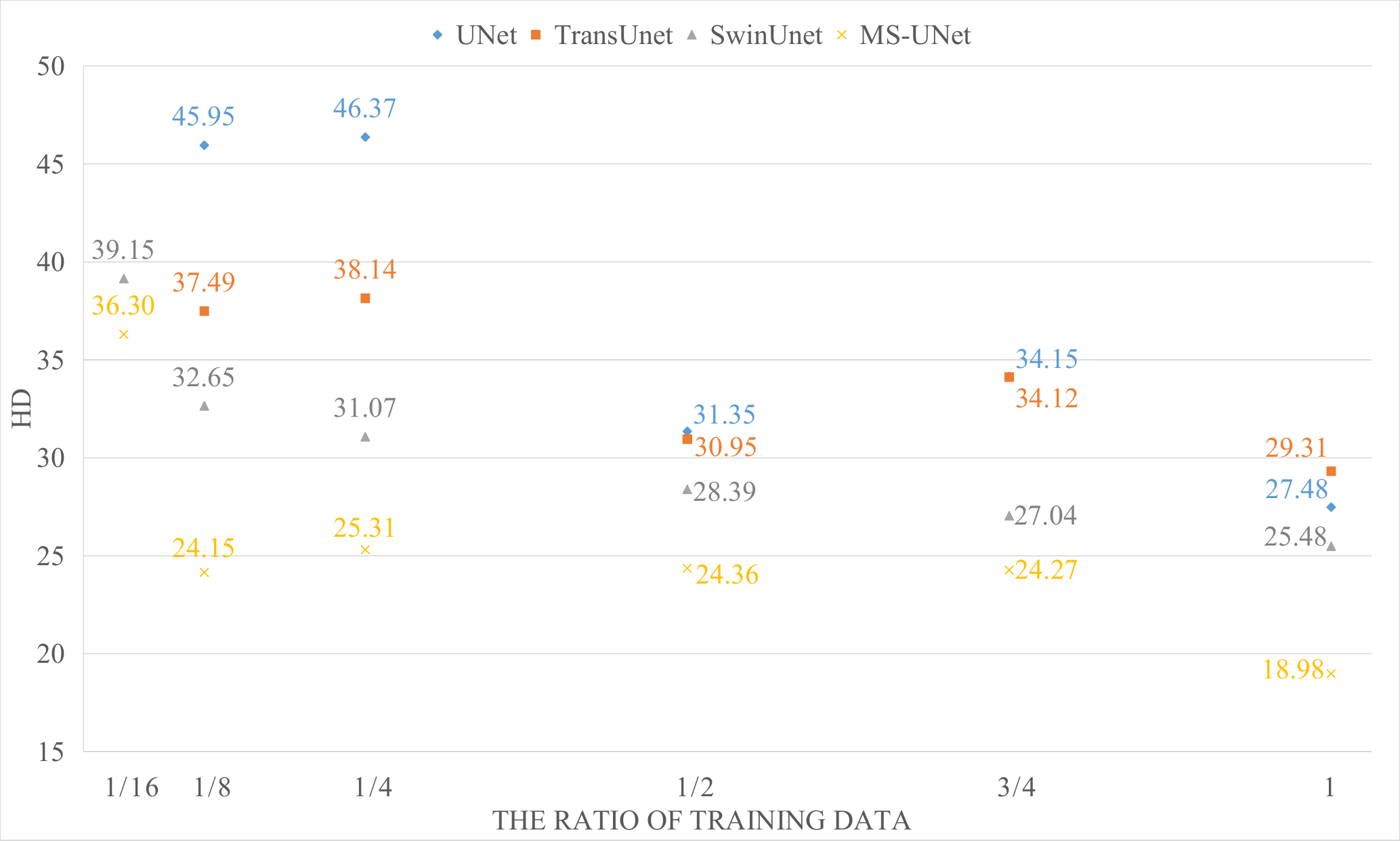}
  \centerline{(d) HD on Synapse Dataset.}\medskip
\end{minipage}

\begin{minipage}{0.48\textwidth}
  \centering
  \includegraphics[width=\linewidth]{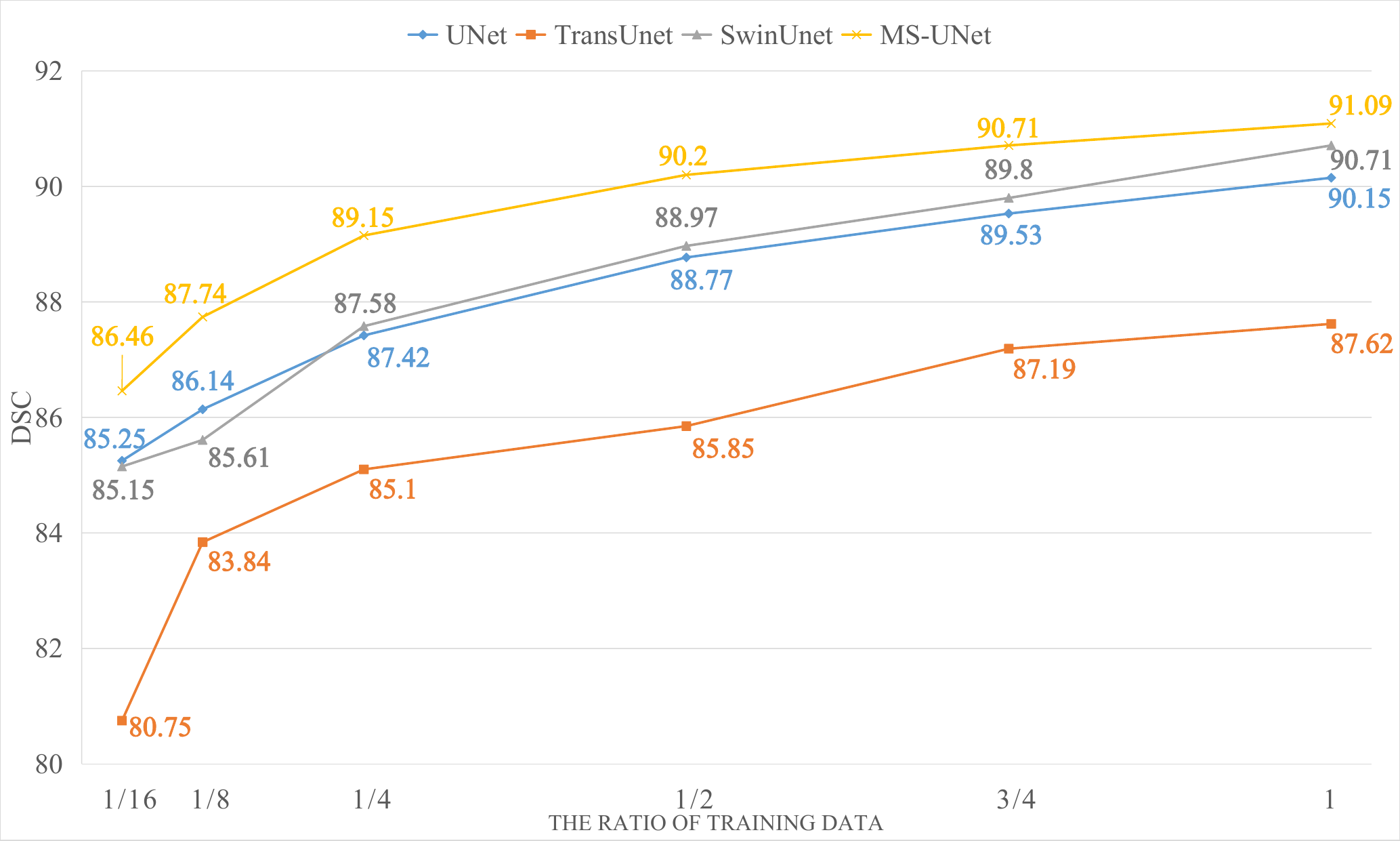}
  \centerline{(b) DSC on ACDC Dataset.}\medskip
\end{minipage}
\hfill
\begin{minipage}{0.48\textwidth}
  \centering
  \includegraphics[width=\linewidth]{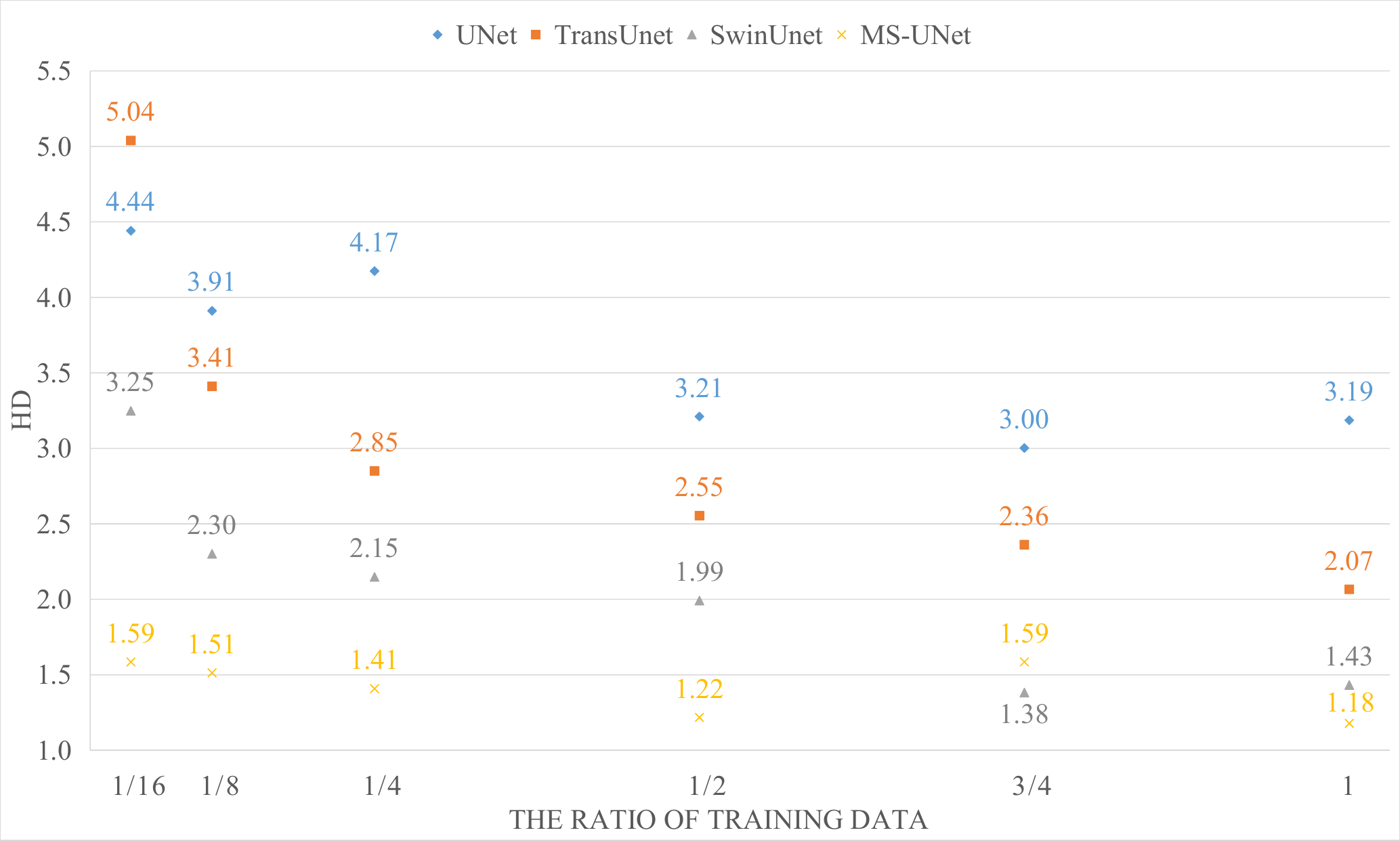}
  \centerline{(e) HD on ACDC Dataset.}\medskip
\end{minipage}

\begin{minipage}{0.48\textwidth}
  \centering
  \includegraphics[width=\linewidth]{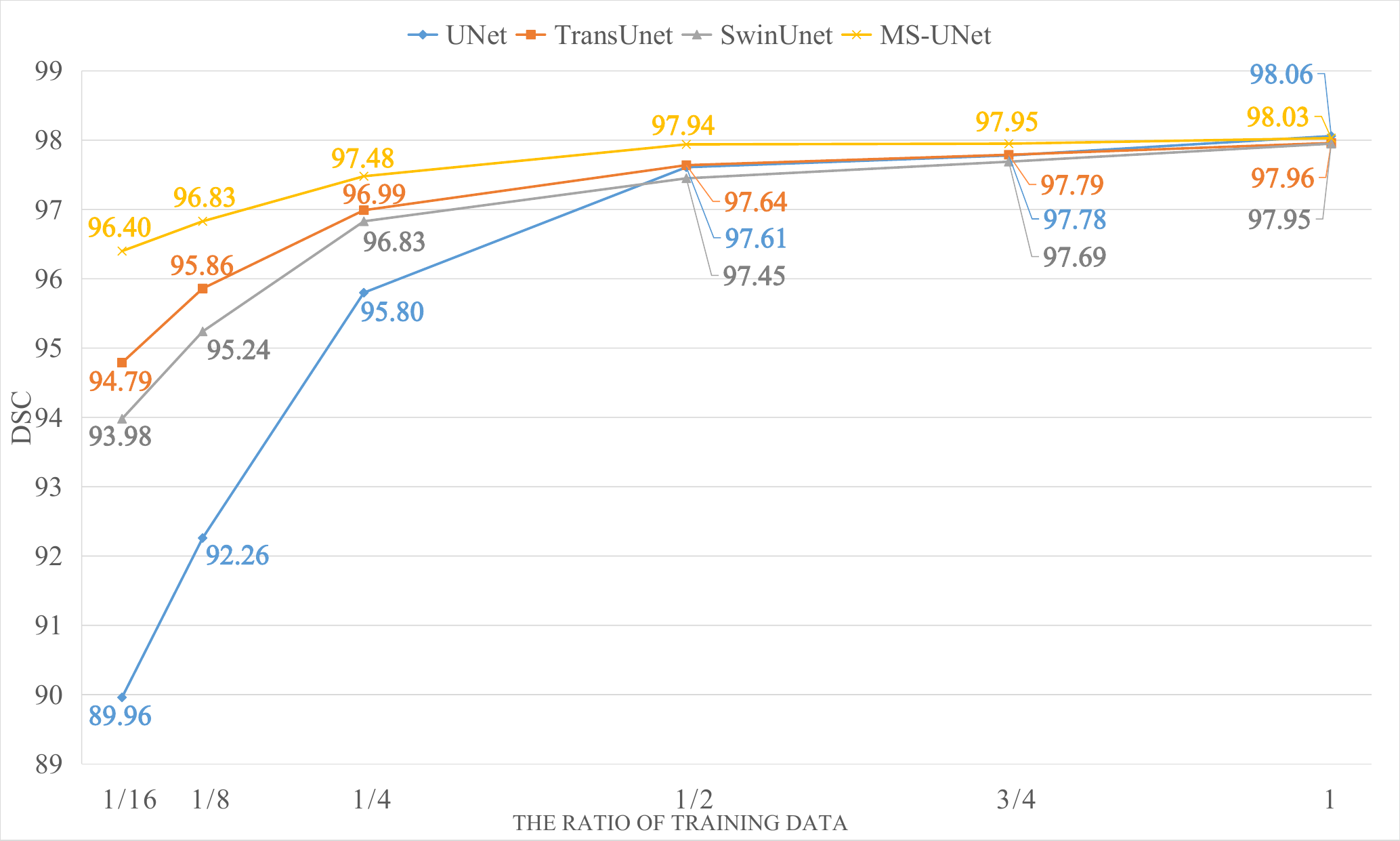}
  \centerline{(c) DSC on JSRT Dataset.}\medskip
\end{minipage}
\hfill
\begin{minipage}{0.48\textwidth}
  \centering
  \includegraphics[width=\linewidth]{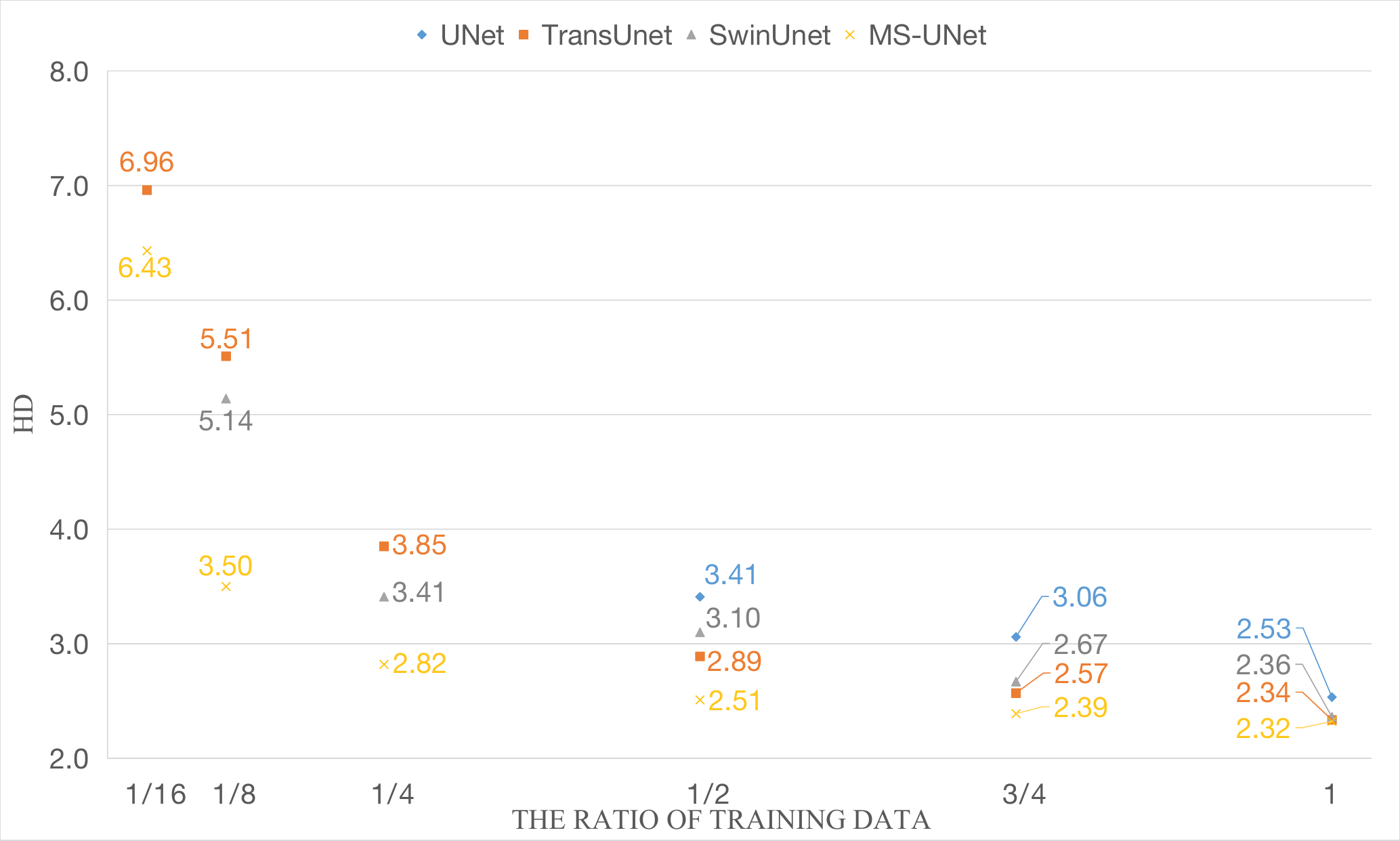}
  \centerline{(f) HD on JSRT Dataset.}\medskip
\end{minipage}
\caption{Results of different methods with different ratios of training data on the Synapse, ACDC, and JSRT datasets.}
\label{fig:Results of segmentation}
\end{figure*}

\section{Experiments}
\label{sec:experiments}
In this section, we will describe the experimental setup, as well as the datasets we have used.
Our proposed MS-UNet is compared with several benchmarks on three public medical image segmentation datasets.
Finally, we have conducted an extensive ablation study on MS-UNet.

\subsection{Datasets}
\subsubsection{Synapse}
Synapse multi-organ segmentation dataset\cite{Syn} includes 30 abdominal CT scans and 3779 axial contrast-enhances abdominal clinical CT images. 
Following the literature\cite{TransUNet, Synapse}, we randomly split the dataset into 18 training sets and 12 testing sets and evaluate our method on 8 abdominal organs (aorta, gallbladder, spleen, left kidney, right kidney, liver, pancreas, spleen, stomach) using the average Dice-Similarity coefficient (DSC) and the average Hausdorff Distance (HD) as evaluation metrics.

\subsubsection{ACDC}
Automated cardiac diagnosis challenge dataset\cite{ACDC} is an open-competition cardiac MRI dataset containing left ventricular (LV), right ventricular (RV), and myocardial (MYO) labels.
The dataset contains 100 samples, which we divided into 70 training samples, 10 validation samples, and 20 test samples for consistency with\cite{MT-UNet}.

\subsubsection{JSRT}
The Japan Society of Radiological Technology dataset\cite{JSRT} provides 247 X-ray images of $1024\times1024$ size, containing lung, heart, and clavicle labels.
In this study, we scale the images to $224\times244$ by interpolation operations and only experiment with lung segmentation.
The JSRT dataset is randomly divided into three independent sets: $172$ for the training data, $25$ for the validation data, and $50$ for the test data.

\subsection{Implementation Details}\label{implementation}
All experiments are conducted on an Nvidia RTX3090 GPU.
The Swin Transformer backbone network in the MS-UNet sub-network has been pre-trained on ImageNet\cite{ImageNet}. 
The input image size in will be set as $224\times224$ and the patch size will be set as $4\time4$. 
During the training process, the default batch size is 24 for all datasets, and we train the model using the SGD optimizer with momentum 0.9, decay $1e{-}4$, learning rate 0.05 for Synapse, 0.0001 for ACDC and JSRT.
We have empirically set the number of epochs for Synapse and ACDC to 250 epochs, and JSRT to 500 epochs.

\subsection{Experimental Results}
Table.\ref{table:Synapse}, Table.\ref{table:ACDC} and Table.\ref{table:JSRT} show the segmentation results of MS-UNet and other up-to-date methods on the Synapse multi-organ CT dataset, ACDC dataset, and JSRT dataset. 
From the results, it could be seen that the proposed MS-UNet exhibits overall better performance than others.
Fig.\ref{fig:Visualization of segmentation1} visualizes segmentation results of different methods on the Synapse multi-organ CT dataset.

In particular, we conduct an experiment to compare the performances of different methods if only using part of the training data.
From Fig.\ref{fig:Results of segmentation}, MS-UNet performs better than other models even with only half of the training data.
This verifies that the proposed model with the multi-scale nested decoder could learn the semantic information of the feature maps from a more multi-dimensional perspective, forming a tighter semantic hierarchy of feature maps between the encoder and decoder parts and achieving better segmentation results even with fewer data.  
It also indicates that the proposed model is more suitable for medical image processing tasks, where the labeled data are rare and costly.

\subsection{Performance on Synapse}
In the Synapse dataset, we chose three open-source typical U-like family models and four results provided in \cite{TransUNet} as baseline and quantitatively compared these results with our proposed MS-UNet model.
The quantitative comparison results are shown in Table.\ref{table:Synapse}.
From the quantitative analysis, it is evident that our proposed MS-UNet model outperforms all the baseline models in terms of the Dice-Similarity Coefficient (DSC) and Hausdorff Distance (HD) metrics.
Across the majority of evaluation indicators, MS-UNet consistently achieves the best results. When compared to the SwinUnet model \cite{SwinUnet}, MS-UNet demonstrates a notable improvement of $1.52\%$ in DSC and $3.74$ in HD.
Moreover, when compared to the other baseline models, MS-UNet exhibits quantitative improvements in the vast majority of organ segmentation metrics.

By incorporating our proposed Edge Loss into MS-UNet, further improvements are achieved. The addition of the Edge Loss results in a $0.44\%$ improvement in DSC and a $0.91$ improvement in HD.
Furthermore, the introduction of the Plug-and-Play Fine-Tuning Denoising Module on the trained MS-UNet model leads to additional enhancements.
This modification yields a $0.94\%$ improvement in DSC and a $2.82$ improvement in HD compared to MS-UNet with Edge Loss.
Notably, our proposed method showcases significant advancements in kidney organ segmentation compared to the other baseline models.
Specifically, the Kidney(L) segmentation demonstrates a remarkable improvement of $4.37\%$ compared to UNet \cite{UNet}, while the Kidney(R) segmentation achieves a substantial enhancement of $5.07\%$ compared to SwinUnet \cite{SwinUnet}.

These quantitative results validate the effectiveness of our proposed MS-UNet model in accurately segmenting organs in the Synapse dataset.
The integration of the Edge Loss and the Plug-and-Play Fine-Tuning Denoising Module further boosts performance, particularly in kidney organ segmentation.
The advancements achieved by our proposed method demonstrate its potential to enhance medical image segmentation tasks and contribute to improved clinical decision-making processes.

\subsection{Performance on ACDC}
The ACDC dataset serves as the second dataset in our experiments, focusing on MRI images.
Similar to the CT images, MRI images are widely used in various medical image processing tasks.
In Table.\ref{table:ACDC}, we present the segmentation results obtained using our proposed method, showcasing incremental improvements over other baseline methods.

The quantitative analysis in Table 2 demonstrates that our approach outperforms the baseline methods in all evaluation metrics for the ACDC dataset.
Specifically, our proposed method achieves a mean DSC of $91.01\% \pm 0.09\%$ and a mean HD of $1.21 \pm 0.05$.
This includes DSC scores of $89.11\%$ for the left ventricular segmentation, $88.50\%$ for the right ventricular segmentation, and $95.41\%$ for the myocardial segmentation.
When incorporating the Edge Loss into the MS-UNet sub-network, we observe further improvements.
The DSC score increases to $91.08\% \pm 0.04\%$, the HD value becomes $1.24 \pm 0.04$, and the individual segmentation categories show DSC scores of $89.10\%$ for the left ventricular, $88.57\%$ for the right ventricular, and $95.58\%$ for the myocardial segmentation.
Similarly to the performance on the Synapse dataset, the integration of our Plug-and-Play Fine-Tuning Denoising Module leads to enhanced segmentation performance on the ACDC dataset.
With the addition of this module, our model achieves a DSC score of $91.26\% \pm 0.04\%$ and an HD value of $1.15 \pm 0.01$.

These results highlight the effectiveness of our proposed method in achieving high-quality segmentation results on the ACDC dataset.
The integration of the Edge Loss and the Plug-and-Play Fine-Tuning Denoising Module further enhances the model's segmentation performance, emphasizing its potential for accurate and reliable medical image segmentation in MRI data.

\begin{table*}
\caption{The average Dice-Similarity coefficient (DSC) and the average Hausdorff Distance (HD) of different methods on the complete Synapse multi-organ CT dataset.}\label{table:Synapse}
\resizebox{1\textwidth}{!}{
\begin{tabular}{c|cc|cccccccc}
\hline
Methods & ${DSC\uparrow}$ & ${HD\downarrow}$ & Aorta & Gallbladder & Kidney(L) & Kidney(R) & Liver & Pancreas & Spleen & Stomach \\
\hline
R50 U-Net\cite{TransUNet} & 74.68 & 36.87 & 87.74 & 63.66 & 80.60 & 78.19 & 93.74 & 56.90 & 85.87 & 74.16\\
R50 Att-UNet\cite{TransUNet} & 75.57 & 36.97 & 55.92 & 63.91 & 79.20 & 72.71 & 93.56 & 49.37 & 87.19 & 74.95\\
Att-UNet\cite{Att-UNet} & 77.77 & 36.02 & \textbf{89.55} & \textbf{68.88} & 77.98 & 71.11 & 93.57 & 58.04 & 87.30 & 75.75\\
U-Net\cite{UNet} & $77.22 \pm 0.41$ & $29.31 \pm 0.85$ & $87.79$ & $65.10$ & $83.26$ & $76.69$ & $94.05$ & $52.50$ & $86.57$ & $71.79$\\
\hline
R50 ViT\cite{TransUNet} & 71.29 & 32.87 & 73.73 & 55.13 & 75.80 & 72.20 & 91.51 & 45.99 & 81.99 & 73.95\\
MT-UNet\cite{MT-UNet} & 78.59 & 26.59 & 87.92 & 64.99 & 81.47 & 77.29 & 93.06 & 59.46 & 87.75 & 76.81\\
TransUnet\cite{TransUNet} & $77.07 \pm 0.34$ & $31.41 \pm 1.27$ & $87.07$ & $61.46$ & $81.47$ & $75.39$ & $94.10$ & $55.83$ & $85.24$ & $77.66$\\
Swin-Unet\cite{SwinUnet} & $78.45 \pm 0.49$ & $25.69 \pm 1.28$ & $85.87$ & $66.62$ & $82.31$ & $78.95$ & $93.87$ & $56.13$ & $89.00$ & $74.84$\\
\hline
\textbf{MS-UNet} & $79.97 \pm 0.20$ & $21.95 \pm 2.11$ & $85.33$ & $67.89$ & $84.86$ & $81.24$ & $94.30$ & $58.24$ & $90.14$ & $77.76$\\
\textbf{MS-UNet+Loss} & $80.41 \pm 0.27$ & $21.04 \pm 0.72$ & $85.25$ & $68.18$ & $86.29$ & $82.30$ & $94.23$ & $59.95$ & $90.08$ & $77.00$\\
\textbf{MS-UNet+Denoising} & \textbf{81.35 $\pm$ 0.02} & \textbf{18.22 $\pm$ 0.05} & $86.33$ & $67.82$ & \textbf{87.63} & \textbf{84.02} & \textbf{94.52} & \textbf{60.77} & \textbf{90.66} & \textbf{79.05}\\
\hline
\end{tabular}
}
\end{table*}

\begin{table*}
\caption{The average Dice-Similarity coefficient (DSC) and the average Hausdorff Distance (HD) for different epoch times to introduce Edge Loss on the complete Synapse multi-organ CT dataset.}\label{table:edge_loss_time}
\resizebox{1\textwidth}{!}{
\begin{tabular}{c|cc|cccccccc}
\hline
Loss Time & ${DSC\uparrow}$ & ${HD\downarrow}$ & Aorta & Gallbladder & Kidney(L) & Kidney(R) & Liver & Pancreas & Spleen & Stomach \\
\hline
Without Loss & $79.97 \pm 0.20$ & $21.95 \pm 2.11$ & \textbf{85.33} & $67.89$ & $84.86$ & $81.24$ & \textbf{94.30} & $58.24$ & $90.14$ & \textbf{77.76}\\
Loss 0 & $80.07 \pm 0.38$ & $22.69 \pm 1.33$ & $85.02$ & \textbf{68.50} & $85.03$ & $81.24$ & $94.06$ & $59.66$ & \textbf{90.22} & $76.86$\\
Loss 50 & \textbf{80.41 $\pm$ 0.27} & \textbf{21.04 $\pm$ 0.72} & $85.02$ & $68.18$ & \textbf{86.29} & \textbf{82.30} & 94.23 & \textbf{59.95} & 90.08 & 77.00\\
Loss 100 & $79.86 \pm 0.12$ & $22.60 \pm 1.51$ & $85.09$ & $67.82$ & $84.85$ & $81.73$ & $94.12$ & $58.53$ & $90.05$ & $76.68$\\
\hline
\end{tabular}
}
\end{table*}

\begin{table*}
\caption{The average Dice-Similarity coefficient (DSC) and the average Hausdorff Distance (HD) for different weights of Edge Loss on the complete Synapse multi-organ CT dataset.}\label{table:edge_loss_omega}
\resizebox{1\textwidth}{!}{
\begin{tabular}{c|cc|cccccccc}
\hline
$\omega$ & ${DSC\uparrow}$ & ${HD\downarrow}$ & Aorta & Gallbladder & Kidney(L) & Kidney(R) & Liver & Pancreas & Spleen & Stomach \\
\hline
0 & $79.97 \pm 0.20$ & $21.95 \pm 2.11$ & 85.33 & $67.89$ & $84.86$ & $81.24$ & \textbf{94.30} & $58.24$ & \textbf{90.14} & \textbf{77.76}\\
0.1 & \textbf{80.41 $\pm$ 0.27} & \textbf{21.04 $\pm$ 0.72} & $85.02$ & \textbf{68.18} & \textbf{86.29} & \textbf{82.30} & 94.23 & \textbf{59.95} & 90.08 & 77.00\\
0.2 & $79.82 \pm 0.35$ & $23.08 \pm 0.88$ & \textbf{84.02} & 67.17 & $85.24$ & $81.44$ & $93.91$ & $59.79$ & 89.63 & $77.41$\\
0.5 & $78.22 \pm 0.26$ & $22.93 \pm 2.10$ & $81.22$ & $65.73$ & $84.96$ & $80.60$ & $93.70$ & $56.80$ & $87.67$ & $75.07$\\
\hline
\end{tabular}
}
\end{table*}

\subsection{Performance on JSRT}
The JSRT dataset serves as the third dataset in our experiments, focusing on X-ray images.
In Table.\ref{table:JSRT}, we present the segmentation results obtained using our proposed method, as well as other baseline methods.

Our MS-UNet sub-network achieves a DSC of $98.0052\% \pm 0.0108\%$ and an HD of $2.3756 \pm 0.0357$. These results demonstrate the superior segmentation performance of our method compared to other Transformer-based methods. The improved network structure of MS-UNet contributes to its exceptional performance on the JSRT dataset.
Furthermore, when incorporating our proposed edge loss, the segmentation performance of MS-UNet becomes comparable to that of UNet, which leverages the advantages of convolutional neural networks on the binary segmentation task with limited data.
With the edge loss, MS-UNet achieves a DSC of $98.0219\% \pm 0.0074\%$ and an HD of $2.3586 \pm 0.0282$.
Lastly, by introducing the denoising module to the MS-UNet sub-network, our approach surpasses UNet and achieves further improvements in segmentation performance.
We report a DSC of $98.0543\% \pm 0.0113\%$ and an HD of $2.2698 \pm 0.0295$.

These results demonstrate the effectiveness of our proposed method in achieving accurate and reliable segmentation results on the JSRT dataset. The integration of the edge loss and the denoising module further enhances the segmentation performance, allowing our approach to outperform UNet and establish itself as a competitive method for medical image segmentation on X-ray images.

\subsection{Ablation Study on Training Data Volume}
To demonstrate the learning efficiency of MS-UNet on few-shot datasets, we conducted an ablation study where we compared the segmentation performance of MS-UNet with other baseline models using varying training data volumes.
Fig.\ref{fig:Results of segmentation} presents the quantitative results of this study.
We randomly selected training data at predetermined rates and trained the models accordingly. The segmentation performance of each trained model was then evaluated using the same test data.
The results clearly indicate that our proposed MS-UNet model outperforms the baseline models on both the Synapse and ACDC datasets, even when only a limited amount of training data is available.

On the Synapse dataset, UNet and TransUnet utilize the full training data and achieve DSCs of $78.23\%$ and $77.93\%$, respectively.
SwinUnet, on the other hand, uses three-quarters of the training data and achieves a DSC of $77.05\%$.
Notably, our proposed MS-UNet model achieves a satisfactory segmentation result of $78.41\%$ DSC using only half of the training data.
This highlights the efficient acquisition of semantic information by MS-UNet even with a smaller training dataset.
Similar observations are made on the ACDC dataset, where MS-UNet consistently exhibits superior segmentation performance compared to the baseline models, even with a reduced amount of training data.
In the case of the JSRT dataset, UNet achieves better segmentation performance than other models when using the complete training data.
This is due to the binary segmentation task being well-suited for convolutional neural network models with simpler structures.
MS-UNet achieves a DSC of $98.03\%$ with all the training data, which is slightly lower than UNet but still better than other Transformer-based baseline models.
As the training data volume decreases, the advantage of a Transformer-based model with pre-trained parameters becomes more evident.
UNet's segmentation performance deteriorates significantly and rapidly with limited training data.
In contrast, our proposed MS-UNet demonstrates remarkable stability in segmentation performance, even with a minimal amount of training data.
In the extreme scenario where only one-sixteenth of the training data is utilized, MS-UNet surpasses other baseline models, achieving a remarkable segmentation performance of $96.40\%$ DSC.
In comparison, TransUnet achieves $94.79\%$ DSC, SwinUnet achieves $93.98\%$ DSC, and UNet achieves $89.96\%$ DSC.

\begin{table}
\caption{The average Dice-Similarity coefficient (DSC) and the average Hausdorff Distance (HD) of different methods on the ACDC dataset.}\label{table:ACDC}
\resizebox{1\linewidth}{!}{
\begin{tabular}{c|cc|ccc}
\hline 
Methods & ${DSC\uparrow}$ & ${HD\downarrow}$ & RV & Myo & LV  \\
\hline
U-Net\cite{UNet} & $89.61 \pm 0.42$ & $2.93 \pm 0.33$ & 85.85 & 87.96 & 95.02 \\
TransUnet\cite{TransUNet} & $87.29 \pm 0.16$ & $2.21 \pm 0.19$ & 83.44 & 84.66 & 93.77 \\
Swin-Unet\cite{SwinUnet} & $90.57 \pm 0.10$ & $1.39 \pm 0.16$ & 88.28 & 88.16 & 95.27 \\
\hline
\textbf{MS-UNet} & 91.01 $\pm$ 0.09 & 1.21 $\pm$ 0.05 & 89.11 & 88.50 & 95.41 \\
\textbf{MS-UNet+Loss} & 91.08 $\pm$ 0.04 & 1.24 $\pm$ 0.04 & 89.10 & 88.57 & \textbf{95.58} \\
\textbf{MS-UNet+Denoising} & \textbf{91.26 $\pm$ 0.04} & \textbf{1.15 $\pm$ 0.01} & \textbf{89.71} & \textbf{88.66} & 95.42 \\
\hline
\end{tabular}
}
\end{table}

\begin{table}
\caption{The average Dice-Similarity coefficient (DSC) and the average Hausdorff Distance (HD) of different methods on the JSRT dataset.}\label{table:JSRT}
\resizebox{1\linewidth}{!}{
\begin{tabular}{c|cc}
\hline 
Methods & ${DSC\uparrow}$ & ${HD\downarrow}$\\
\hline
U-Net\cite{UNet} & $98.0234 \pm 0.0271$ & $2.5822 \pm 0.0876$ \\
TransUnet\cite{TransUNet} & $97.9181 \pm 0.0422$ & $2.4609 \pm 0.0646$ \\
Swin-Unet\cite{SwinUnet} & $97.9218 \pm 0.0145$ & $2.4662 \pm 0.0426$ \\
\hline
\textbf{MS-UNet} & 98.0052 $\pm$ 0.0108 & 2.3756 $\pm$ 0.0357 \\
\textbf{MS-UNet+Loss} & 98.0219 $\pm$ 0.0074 & 2.3586 $\pm$ 0.0282 \\
\textbf{MS-UNet+Denoising} & \textbf{98.0543 $\pm$ 0.0113} & 2.2698 $\pm$ 0.0295\\
\hline
\end{tabular}
}
\end{table}

These results validate the effectiveness of the proposed MS-UNet model in capturing semantic information through its multi-scale nested decoder.
The robust semantic hierarchy established between the encoder and decoder components enables MS-UNet to achieve superior segmentation performance, even when confronted with limited data availability.
The findings also highlight the practical significance of MS-UNet in medical image processing tasks, where labeled data are often scarce and expensive to acquire.
By effectively utilizing available data and demonstrating stability in segmentation performance, MS-UNet offers a valuable solution for addressing the challenges associated with limited labeled data in medical imaging applications.

\subsection{Ablation Study on Network Architectures}
In this part, we explore the impact of the number and location of multi-scale nested decoders on the segmentation performance of MS-UNet by changing them gradually. 
From the experimental results in Table.\ref{table:multi_level_nested_blocks}, even when adding only one multi-scale nested block in a shallow layer of the network, the DSC of the model improves from $79.23\%$ to $80.12\%$, and the number of parameters has increased by only $1\%$. 
This suggests that the introduction of multi-scale nested decoders enhances the model's ability to capture and utilize semantic information, leading to improved segmentation results. 
Moreover, as the number of multi-scale nested blocks increases, the segmentation performance of the model continues to improve without overfitting.
This observation is particularly important because it demonstrates that the benefits of the multi-scale nested decoder structure are not limited to a specific number or location of the blocks.
The model could effectively leverage multiple nested decoders to bring the feature maps between the encoder and decoder closer in a semantic sense, resulting in better segmentation performance.

These findings further support the notion that the structure of MS-UNet with multi-scale nested decoders enables the network to capture and utilize semantic information more effectively.
By semantically connecting the encoder and decoder components, the model could better exploit the hierarchical relationships within the feature maps, leading to improved segmentation performance.

\begin{figure}
\centering
\includegraphics[width=1\linewidth]{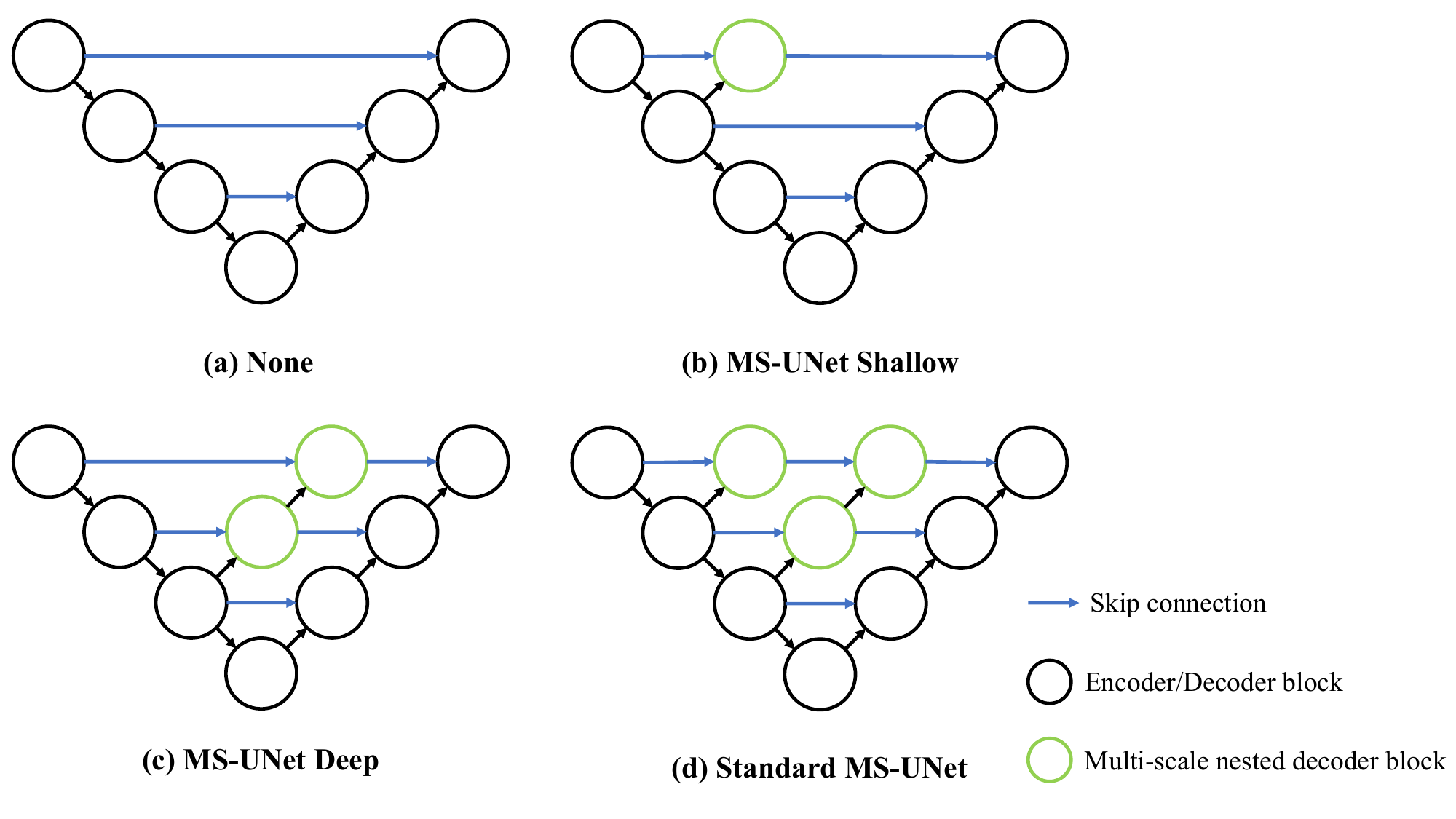}
\caption{The architecture of the network with different numbers and locations of the multi-scale nested blocks.} 
\label{fig:Structure_dif_blocks}
\end{figure}

\begin{table}
\caption{The best Dice-Similarity coefficient (DSC) and Hausdorff Distance (HD) of different MS-UNet structures with Mutil-scale Nested Decoder (MND) blocks on the Synapse multi-organ CT dataset.}\label{table:multi_level_nested_blocks}
\resizebox{1\linewidth}{!}{
\begin{tabular}{c|c|cc}
\hline
Structure & Parmas & ${DSC\uparrow}$ & ${HD\downarrow}$\\
\hline
None & 27.17 M & 79.23 & 25.69\\
MS-UNet Structure2 & 27.49 M & 80.13 & 20.95 \\
MS-UNet Structure1& 28.75 M & 80.16 & 21.79\\
\hline
\textbf{Standard MS-UNet} & 29.06 M & \textbf{80.44} & \textbf{18.97}\\
\hline
\end{tabular}
}
\end{table}

\subsection{Ablation Study on Edge Loss}
As stated in Eq. (\ref{Loss}), we assigned a fixed value of 0.5 to both $\omega_1$ and $\omega_2$.
This decision is made to emphasize the significance of the Cross-Entropy loss and the Dice-Similarity Coefficient loss at the pixel level, recognizing their importance in achieving accurate semantic segmentation for our task.
To ascertain the optimal weighting of $\omega_3$, we conducted ablation experiments, while the quantitative results are presented in Table.\ref{table:edge_loss_omega}, and Fig.\ref{fig:Visualization of segmentation2} visualizes segmentation results of different architectures of our proposes on the Synapse multi-organ CT dataset.
By incorporating the Edge Loss with a weight ($\omega3$) of 0.1 into the MS-UNet sub-network, the segmentation performance of the model is notably enhanced.
The DSC improves from an initial value of $79.97\% \pm 0.20\%$ to $80.41\% \pm 0.27\%$, and the HD decreases from $21.95 \pm 2.11$ to $21.04 \pm 0.72$.
This indicates that the Edge Loss contributes to improving the segmentation accuracy of the model.
Specifically, the segmentation performance improves from an initial DSC of $79.97\% \pm 0.20\%$ and HD of $21.95 \pm 2.11$ to a DSC of $80.41\% \pm 0.27\%$ and HD of $21.04 \pm 0.72$.
However, it is important to note that as $\omega_3$ increases beyond a certain point, the segmentation performance of the model starts to decline. 
In fact, it could even fall below the performance achieved by the MS-UNet sub-network without the inclusion of the Edge Loss. 
This suggests that the Cross-Entropy Loss and DSC Loss, which represent the overall segmentation region loss, still play a dominant role in model training.
Finding an optimal weight for the Edge Loss is crucial to maintain and improve the overall segmentation performance

Furthermore, the timing of introducing the Edge Loss during the training process also impacts the segmentation performance.
In order to investigate the optimal timing for introducing the Edge Loss in the training process, we conducted a series of ablation experiments using the parameter settings of Sec.\ref{implementation} as the benchmark. 
Specifically, we introduced the Edge Loss at different epochs, namely, the 0th, 50th, and 100th epochs. 
This experimental setup allowed us to assess the impact of incorporating the Edge Loss at various stages of training.
The results of the experiment are shown in Table.\ref{table:edge_loss_time}.
According to the Table.\ref{table:edge_loss_time}, it is observed that introducing the Edge loss from the beginning of the training leads to a slight improvement in the segmentation performance of the model, increasing the DSC from $79.97\% \pm 0.20\%$ to $80.07\% \pm 0.38\%$, but decreasing the HD from $21.95 \pm 2.11$ to $22.69 \pm 1.33$.
When we introduce the edge loss at the 50th epoch, the model achieves the best segmentation performance with a DSC of $80.41\% \pm 0.27\%$ and HD of $21.04 \pm 0.72$.
Interestingly, delaying the introduction of the edge loss beyond the 50th epoch leads to a decrease in segmentation performance, ultimately reaching a level similar to that of the model without the inclusion of the edge loss. 
This emphasizes the importance of introducing the Edge Loss at an appropriate stage during training to achieve optimal segmentation performance.

Overall, these experiments highlight the significance of the Cross-Entropy Loss and DSC Loss for effective and accurate segmentation results, while the Edge Loss serves as a supplementary supervisory signal.
Finding the right weight for the Edge Loss and determining the optimal timing for its introduction are critical steps in maximizing the segmentation performance of the MS-UNet model.

\subsection{Discussion and Future Work}
In this work, we apply the proposed MS-UNet, Edge loss, and Denoising module for medical image segmentation tasks.
According to the results in Fig.\ref{fig:Results of segmentation}, our model achieves overall better results relative to other models, even with a small amount/part of training data.
We need to emphasize that in the extreme case where only a very small amount of training data is available, the improvement brought by the proposed method is huge.
Unlike other computer vision tasks, in medical image tasks, it is usually difficult to obtain labeled medical images since it is usually costly.
From this perspective, the proposed MS-UNet has a relatively smaller data scale requirement and has great potential for different medical image processing tasks.
At the same time, to further enhance the segmentation performance of the model, we propose Edge loss and a Plug-and-play Fine-tuning Denoising module, which act on the training phase and the completed phase of the model respectively.
As depicted in Table.\ref{table:Synapse}\ref{table:ACDC}\ref{table:JSRT}, the results demonstrate the effectiveness of our proposed Edge loss and Denoising module in enhancing the segmentation performance of the model. 
In future work, we plan to undertake further optimization of these modules to effectiveness in improving segmentation performance. 
Additionally, we intend to explore the applicability of these modules on other models, extending their benefits to a broader range of segmentation tasks. 
For better comparison with\cite{SwinUnet, TransUNet, MT-UNet}, we only use MS-UNet for 2D segmentation tasks.
We are planning to improve MS-UNet for 3D segmentation tasks in the future.
In addition, with the popularity of \cite{MoCo, MAE}, We will further investigate how to pre-train the Transformer-based model on unlabelled medical images by a self-supervised approach and combine it with our proposed methods.

\section{Conclusion}
In this work, a novel Transformer-based U-shaped medical image segmentation network called MS-UNet has been proposed.
MS-UNet addresses the challenge of capturing semantic information between the encoder and decoder components by introducing a multi-scale nested decoder structure.
This design enables a tighter semantic hierarchy and improves the stability and generalization of the model.
Experimental results have demonstrated the excellent performance and efficient feature learning ability of MS-UNet.
Notably, MS-UNet outperforms other U-Net family models even in scenarios with a very limited amount of training data.
This is a significant contribution, considering the scarcity and costliness of labeled medical images.
Additionally, this work introduces a novel edge loss and a companion edge label generation method, which collectively enhance the segmentation performance of MS-UNet.
The edge loss emphasizes the importance of edge information, further improving the accuracy of the segmentation results.
Furthermore, a plug-and-play fine-tuning denoising module is proposed, which could be utilized not only with MS-UNet but also with other segmentation models.
This module effectively reduces noise in the completed phase of the model, leading to improved segmentation performance.

Overall, this work has great potential for medical image processing tasks where labeled images are rare and expensive.
The contributions of MS-UNet, the edge loss, and the denoising module provide valuable tools for accurate and efficient segmentation in real-world medical applications.

\label{sec:refs}

\bibliographystyle{IEEEbib}
\bibliography{refs}

\end{document}